\documentstyle[11pt,aaspp4,epsf]{article}

\begin{document}

\title{EVIDENCE IN VIRGO FOR THE UNIVERSAL DARK MATTER HALO}
\author{Dean E.~McLaughlin\altaffilmark{1}}
\affil{Department of Astronomy, 601 Campbell Hall\\
University of California, Berkeley, CA 94720-3411\\
dean@crabneb.berkeley.edu}

\altaffiltext{1}{Hubble Fellow}

\lefthead{{\sc McLAUGHLIN}}
\righthead{THE DARK MATTER HALO OF VIRGO}

\begin{abstract}

A model is constructed for the mass and dynamics of M87 and the Virgo Cluster.
Existing surface photometry of the galaxy, mass estimates from X-ray
observations of the hot intracluster gas, and the velocity dispersions of
early-type Virgo galaxies, all are used to constrain the run of dark matter
density over $0\le r\la 2$ Mpc in the cluster. The ``universal'' halo
advocated by Navarro, Frenk, \& White, $\rho_{\rm dm}\propto(r/r_s)^{-1}
(1+r/r_s)^{-2}$, provides an excellent description of the combined data, as
does a Hernquist profile with $\rho_{\rm dm}\propto(r/r_s)^{-1}(1+r/r_s)^{-3}$.
These models are favored over isothermal spheres, and their central structure
is preferred to density cusps either much stronger or much weaker
than $r^{-1}$. The galaxies and gas in the cluster trace its
total mass distribution ($\rho_{\rm gal}\propto\rho_{\rm gas}\propto
\rho_{\rm dm}$), the galaxies' velocity ellipsoid is close to isotropic, and
the gas temperature follows the virial temperature profile of the dark halo.
The virial radius and mass and the intracluster gas fraction of Virgo are
evaluated.

\end{abstract}


\section{Introduction}

Some recent simulations of structure formation in CDM universes have suggested
that dark matter halos generally form with a ``universal'' density profile, in
which $\rho_{\rm dm}\propto(r/r_s)^{-1}(1+r/r_s)^{-2}$
(\markcite{nav97}Navarro, Frenk, \& White 1997, hereafter NFW). However, this
claim has been questioned by other numerical work (e.g.,
\markcite{ann96}Anninos \& Norman 1996; \markcite{kra98}Kravtsov et al.~1998);
in particular, \markcite{moo98}Moore et al.~(1998) find instead that
$\rho_{\rm dm}\rightarrow r^{-1.4}$ as $r\rightarrow0$ in a set of highly
resolved N-body halos. The little {\it observational} evidence that has been
brought to bear on this issue is also ambiguous; e.g., the basic form of the
\markcite{\nav97}NFW halo appears to apply in CNOC clusters of galaxies
(\markcite{car97}Carlberg et al.~1997), but not in the dwarf galaxy DDO154
(\markcite{bur97}Burkert \& Silk 1997). This {\it Letter} tests the viability
of the proposed universal halo, among other models for the dark matter
distribution, in the nearby Virgo Cluster ($D=15$ Mpc; \markcite{pie94}Pierce
et al.~1994).

The X-ray emission from Virgo, which is centered on the cD galaxy M87, has
been used many times to constrain mass models for the cluster core
(\markcite{nul95}Nulsen \& B\"ohringer 1995, and references therein). Here,
an attempt is made to supplement this standard analysis with larger-scale
constraints on $\rho_{\rm dm}(r)$ from the spatial distribution and dynamics
of the galaxies in Virgo. To do so, the infalling late-type galaxies that are
responsible for much of the cluster's irregular optical structure are
excluded from analysis. Instead, reference will be made only to the {\it early}
Hubble types in Virgo, which are spatially concentrated (about a point
$\sim1\arcdeg$ NW of M87) with a roughly Gaussian velocity distribution
(centered near the velocity of M87), and which therefore suggest the existence
of a relatively smooth and relaxed underlying mass distribution (see
\markcite{bts87}Binggeli, Tammann, \& Sandage 1987). In what follows, this is
essentially adopted as a postulate. It is further assumed that Virgo is
spherically symmetric and centered on M87. This is a reasonable first-order
characterization of the X-ray structure around the galaxy, and will serve as
a coarse-average approximation to the overall cluster structure. As part
of this, the apparent offset between M87 and the centroid of the early-type
galaxy isopleths is taken {\it not} to reflect a large perturbation in the
fundamental mass distribution, and is ignored. While imperfect, these
simplifications ultimately admit a model for Virgo that is of interest both
for its basic structure---the central scaling $\rho_{\rm dm}\propto r^{-1}$ of
\markcite{\nav97}NFW is indicated, and is favored clearly over the steeper
cusps of either \markcite{moo98}Moore et al.~(1998) or a singular isothermal
sphere---and for the relations it suggests between the galaxies, gas, and dark
matter in this cluster.

\section{Data and Halo Models}

A significant contribution to the total mass on the smallest spatial scales in
Virgo comes from the stars in M87 itself. With the {\it stellar} (core)
mass-to-light ratio of this galaxy known (\markcite{vdm91}van der Marel 1991),
and taken to be independent of radius, the mass density profile $\rho_{\rm
M87}(r)$ follows from the $B$-band surface photometry of \markcite{dvn78}de
Vaucouleurs \& Nieto (1978), which extends to $R\simeq100$ kpc. (Here and
throughout, three-dimensional radii are denoted by $r$, and projected
radii by $R$.) Specifically, fitting $\mu_B$ with density models from the
family discussed by \markcite{deh93}Dehnen (1993) and \markcite{tre94}Tremaine
et al.~(1994) yields the profile defined in Table \ref{tab1} (for $D=15$ Mpc,
and with $A_B=0.09$ mag of extinction [\markcite{bur84}Burstein \& Heiles
1984] taken into account). Note that this model closely approximates a
standard $R^{1/4}$ law in projection (see \markcite{deh93}Dehnen 1993).

Moving outwards, density and temperature profiles for the hot gas within $10
\la r\la 200$ kpc of M87 have been obtained from {\it ROSAT} PSPC observations
by \markcite{nul95}Nulsen \& B\"ohringer (1995, hereafter NB95). Their
analysis assumes homogeneity, or a single-phase medium, which could be a
potential source of error {\it if} a cooling flow has produced a multiphase
gas in Virgo; however, this possibility appears to be of concern only at
radii near the low end of the range in the {\it ROSAT} data (viz.~$r\la 15$
kpc: \markcite{can82}Canizares et al.~1982; \markcite{tsa94}Tsai 1994).
Thus, use is made here of the {\it model-independent} gravitating mass
profile, $M_{\rm tot}(r)$, which \markcite{nul95}NB95 derive by applying the
standard assumption of hydrostatic equilibrium to their $\rho_{\rm gas}(r)$
and $T(r)$ measurements.

Still further out, the (azimuthally averaged) surface density of dwarf
ellipticals in Virgo is given by \markcite{bts87}Binggeli et al.~(1987), who
find $N_{\rm gal}(R)\propto[1+(R/410\,{\rm kpc})^2]^{-1}$ for
$100\,{\rm kpc}\la R\la 1.5\,{\rm Mpc}$. (The distribution of E/S0 systems,
though somewhat more irregular in detail, is also consistent with this average
profile.) The offset between the X-ray and optical centroids of Virgo is small
enough---about half a core radius, and not much larger than M87 itself---that
the same $N_{\rm gal}(R)$ applies to galaxy counts taken in rings centered on
either point. \markcite{bts87}Binggeli et al.'s
parametrization then implies a three-dimensional $n_{\rm gal}(r)\propto
[1+(r/410\,{\rm kpc})^2]^{-3/2}$. The galaxy data {\it do not
require} this constant-density core, however; they are equally well fit by 
projections of either a \markcite{her90}Hernquist (1990) profile, $n_{\rm gal}
(r)\propto(r/1.1\,{\rm Mpc})^{-1}(1+r/1.1\,{\rm Mpc})^{-3}$, or an
\markcite{nav97}NFW profile, $n_{\rm gal}\propto(r/560\,{\rm kpc})^{-1}(1+
r/560\,{\rm kpc})^{-2}$ (cf.~Fig.~\ref{fig2} below).

Finally, \markcite{gir96}Girardi et al.~(1996) use the velocities of
180 early-type Virgo galaxies to measure the {\it aperture} velocity
dispersion, $\sigma_{\rm ap}$---the r.m.s.~line-of-sight velocity for all
objects within a given distance from M87---as a function of $R$ from
$\simeq200$ kpc to 2 Mpc. Essentially the same $\sigma_{\rm ap}(R)$ profile is
obtained with circular apertures taken about the optical center of Virgo
instead (cf.~\markcite{fad96}Fadda et al.~1996).

Given these data, a parametric model for the dark matter halo may be specified
as $\rho_{\rm dm}(r)\equiv K\,f(r/r_s)$, with $f$ a dimensionless form
function. It is then convenient to refer the halo's normalization and scale
to the observed stellar quantities in M87 (Table \ref{tab1}). Thus, with
$D\equiv\pi a^3 K/\Upsilon_B L_B$ and $\eta\equiv r_s/a$, the mass interior to
any radius is
\begin{equation}
M_{\rm tot}(r) = \Upsilon_B L_B \left[\left({{r/a}\over{1+r/a}}\right)^
{3-\gamma} + 4D \eta^3 \int_0^{r/r_s} f(u) u^2\,du \right]\ ,
\label{eq:2}
\end{equation}
in which $D$ and $\eta$ are fixed by fitting to the observed mass profile at
$r\la 200$ kpc. If a good fit is achieved, no free parameters are left in the
model, and it remains only to check its extrapolation to larger radii against
the dynamics of the early-type galaxies. To this end, the observed $n_{\rm gal}
(r)$ (from any of the functional fits mentioned above) and the model $M_{\rm
tot}(r)$ are used to solve the spherical Jeans equation for the radial velocity
dispersion of the galaxies:
$$n_{\rm gal}\sigma_r^2(r)=\exp\left(-\int {{2\beta_{\rm gal}}\over{r}}\,dr
\right)\ \times\ \qquad\qquad\qquad\qquad\quad$$
\begin{equation}
\qquad\qquad
\left[\int_{r}^\infty n_{\rm gal}{{GM_{\rm tot}}\over{x^2}}
\exp\left(\int{{2\beta_{\rm gal}}\over{x}}\,dx\right)\,dx\right]\ ,
\label{eq:3}
\end{equation}
where $\beta_{\rm gal}(r)\equiv1-\sigma_{\theta}^2/\sigma_r^2$ is the usual
measure of orbital anisotropy. The dispersion along any single line of sight
then follows from the projection
\begin{equation}
N_{\rm gal}\sigma_p^2(R) = 2\,\int_{R}^\infty n_{\rm gal}\sigma_r^2(r)
\left(1-\beta_{\rm gal}\,{{R^2}\over{r^2}}\right)\,{{r\,dr}\over
{\sqrt{r^2-R^2}}}\ ,
\label{eq:4}
\end{equation}
with $N_{\rm gal}(R)=2\int_R^\infty n_{\rm gal}\,(r^2-R^2)^{-1/2} r\,dr$.
Finally,
\begin{equation}
\sigma_{\rm ap}^2(R) =
\left(\int_0^R N_{\rm gal}\sigma_p^2(x)\,x\,dx\right)
\left(\int_0^R N_{\rm gal}(x)\,x\,dx\right)^{-1}
\label{eq:5}
\end{equation}
is compared with the observed aperture velocity dispersion profile at $R\ga
200$ kpc. If this comparison is unsatisfactory, but the basic assumption of
dynamical equilibrium is retained, then the underlying form of $\rho_{\rm dm}
(r)$ may be rejected. A potential complication is that the orbital anisotropy
of the galaxies is not known a priori, and a given $f(r/r_s)$ can safely be
discarded only if it cannot reproduce the empirical $\sigma_{\rm ap}(R)$ for
{\it any} assumed $\beta_{\rm gal}(r)$. Fortunately, the Virgo data prove to
be quite consistent with a simple $\beta_{\rm gal}\equiv0$, i.e., isotropic
orbits.

\section{Results}

The field of possible halo profiles can be narrowed somewhat by first
requiring consistency with a small subset of the data: from
\markcite{nul95}NB95 and \markcite{gir96}Girardi et al.~(1996),
\begin{eqnarray}
M_{\rm tot}(r\la30\,{\rm kpc}) & = & (1.6^{+0.3}_{-0.7})\times10^{12}M_\odot
\nonumber \\
M_{\rm tot}(r\la150\,{\rm kpc}) & = & (1.9\pm0.8) \times 10^{13} M_\odot
\label{eq:1} \\
\sigma_{\rm ap}(R\le2\,{\rm Mpc}) & = & 640^{+85}_{-65}\ {\rm km s}^{-1}\ ,
\nonumber
\end{eqnarray}
where all uncertainties are $\simeq2$-$\sigma$ limits. Figure \ref{fig1}
illustrates tests of four different form functions against these three
constraints only: in panel (a), $f(r/r_s)=[1+(r/r_s)^2]^{-1}$ (an
isothermal sphere with a core; cf.~\markcite{nul95}NB95); in (b), $f(r/r_s)=
(r/r_s)^{-0.5}(1+r/r_s)^{-2.5}$; in panel (c), $f(r/r_s)=(r/r_s)^{-1}(1+r/r_s)
^{-2}$ (the model of \markcite{nav97}NFW); and in (d), $f(r/r_s)=(r/r_s)^{-1.5}
(1+r/r_s)^{-1.5}$ (as in the N-body simulations of \markcite{moo98}Moore et
al.~1998). The different line types in each panel contain combinations
of $D$ and $\eta$ that are consistent (for $\beta_{\rm gal}\equiv0$) with the
2-$\sigma$ limits on the {\it individual} data points in equation (\ref{eq:1});
the various dots identify models which yield the best
estimates for both members of different data {\it pairs}. A minimally
self-consistent model for $\rho_{\rm dm}$ must draw its normalization and
scale length from the intersection of the three bands in the appropriate panel
of Fig.~\ref{fig1}, and ideally the three dots should coincide at a unique
$(\eta,D)$.

Panel (d) of Fig.~\ref{fig1} shows that a central cusp as steep as
$r^{-1.5}$ (\markcite{moo98}Moore et al.~1998) cannot easily account for the
observed $M_{\rm tot}$ inside both 30 kpc and 150 kpc: the implied $M_{\rm
dm}\propto r^{1.5}$ at small $r$ is {\it shallower} than what is inferred from
the observed excess of $M_{\rm tot}(r)$ over the stellar $M_{\rm M87}(r)$. This
is even more of a problem for the singular isothermal sphere ($\rho_{\rm dm}
\propto r^{-2}$ and $M_{\rm dm}\propto r$), which is altogether inconsistent
with the X-ray masses (see also \markcite{nul95}NB95). On the other hand, the
mass constraints at $r\la 200$ kpc are satisfied by any of the halos with
central $\rho_{\rm dm}\propto r^{-1}$ or shallower; but the weak cusps in
panels (a) and (b) give the best agreement for scale radii $r_s$ that are so
small [to limit a {\it steeply} rising $M_{\rm dm}(r)$] as to imply total
halo masses which are too low to support the velocity dispersion of the Virgo
galaxies on large spatial scales.\footnotemark
\footnotetext{This conclusion is insensitive to the assumed form of $\rho_{\rm
dm}$ as $r\rightarrow\infty$. Also, while some shallow-cusp models with radial
anisotropy ($0<\beta_{\rm gal}\le 1$) can reproduce the observed $\sigma_{\rm
ap}$ at $R=2$ Mpc, they are inconsistent with the full {\it profile} at smaller
radii.}
Thus, the ``universal'' halo of \markcite{nav97}NFW, with its central
$\rho_{\rm dm}\propto r^{-1}$ and with $\beta_{\rm gal}\equiv0$, emerges as
the best candidate for a self-consistent description of the Virgo
Cluster [panel (c)].

This simple analysis is limited in two respects. (1) At very
small $r\la 10$--15 kpc, where $M_{\rm M87}(r)\gg M_{\rm dm}(r)$, the halo
could depart from $\rho_{\rm dm}\propto r^{-1}$ and not conflict with any
observations. And (2) little can be said about the behavior of
$\rho_{\rm dm}$ in the limit $r\rightarrow\infty$; for example, the same
procedure that leads to Fig.~\ref{fig1} shows that the profile of
\markcite{her90}Hernquist (1990), $\rho_{\rm dm}\propto(r/r_s)^{-1}(1+r/r_s)^
{-3}$ (as suggested by the early simulations of \markcite{dub91}Dubinski \&
Carlberg 1991), is just as acceptable as the \markcite{nav97}NFW
$\rho_{\rm dm}\propto (r/r_s)^{-1}(1+r/r_s)^{-2}$ in Virgo. The best fits of
these two models have different {\it parameters} (given in Table \ref{tab1}),
but show no real {\it physical} differences on $\la 2$ Mpc scales.

Figure \ref{fig2} compares the full sets of data discussed in \S2 with models
using the best-fit \markcite{nav97}NFW and \markcite{her90}Hernquist halos.
The top panel shows the best X-ray estimate and 95\% confidence limits
for $M_{\rm tot}(r)$ (dashed line and bold curves), along with results from
optical spectroscopy of the stars
in M87 (\markcite{sar78}Sargent et al.~1978;
filled triangles) and from radial velocities of its globular clusters
(\markcite{coh97}Cohen \& Ryzhov 1997; open triangles). These
latter sets of data independently corroborate the models. (The globular
cluster dynamics specifically will be discussed in a future paper.)
The middle panel of Fig.~\ref{fig2} then compares the surface density profile
of early-type galaxies with that of each fit to
the dark matter halo. This shows---the analysis {\it did not assume}---that
$n_{\rm gal}\propto\rho_{\rm dm}$ in Virgo. This fact is used to compute
aperture dispersion profiles that, for $\beta_{\rm gal}\equiv0$, show
excellent agreement with the data in the bottom of Fig.~\ref{fig2}. It also
allows the uncertainties on $\eta$ in Table \ref{tab1} to be estimated by
fitting the projection of $\rho_{\rm dm}(r)$ to $N_{\rm gal}(R)$; the
limits on $D$ then follow from plots like panel (c) of Fig.~\ref{fig1}.

In either of these models, the virial radius of the dark matter halo---roughly,
that within which its mean density is 200 times the critical density for
closure---is $r_{200}=1.55\pm0.06$ Mpc (for $H_0=70$ km s$^{-1}$ Mpc$^{-1}$),
corresponding to a mass of $M_{\rm dm}(r_{200})=(4.2\pm 0.5)\times10^{14}\,
M_\odot$. This emphasizes that the dark matter (and the gas) around M87
is associated---by construction---with the whole of Virgo; and this weakens
previous arguments (\markcite{bts87}Binggeli et al.~1987) which suggested that
the cluster could not be relaxed.

Finally, Fig.~\ref{fig3} shows that the hot gas within 200 kpc of M87 directly
traces the distribution of dark matter (and galaxies) in Virgo:
$\rho_{\rm gas}/\rho_{\rm dm}\simeq0.035\,(D/15\,{\rm Mpc})^{1.5}$. If this is
assumed to hold throughout the entire cluster, then the similarity of its total
stellar and gas masses (\markcite{arn92}Arnaud et al.~1992) implies a global
baryon fraction of $\sim7\%$, roughly in line with other low-temperature
clusters (\markcite{arn98}Arnaud \& Evrard 1998). The bottom of Fig.~\ref{fig3}
also shows a close agreement between the observed gas temperatures and the
virial temperature, $kT(r)=0.6 m_H\sigma_r^2(r)$, of the dark matter halo.
Thus, simple virialization of gas that traces the dark matter in an
$r^{-1}$ halo suffices to explain the density and temperature structure of the
intracluster medium at the center of Virgo. This suggests---if an
\markcite{nav97}NFW-type halo truly is universal, and if the proportionality
$\rho_{\rm gas}\propto\rho_{\rm dm}$ is typical---that central density cusps
and temperature drops in the gas of X-ray clusters {\it may not, by themselves,
prove the existence of cooling flows}. That said, however,
radiative cooling times are unequivocally short in the core of Virgo
$(t_{\rm cool}\la 10^{10}$ yr for $r\la 70$ kpc); and it remains to be seen
whether a single-phase gas virialized in an \markcite{nav97}NFW halo can
account for the low-energy X-ray line fluxes observed there, or if these still
require mass drop-out and the development of a multiphase medium at small radii
in a cooling flow (cf.~Tsai \markcite{tsa94}1994).

Of course, Virgo by itself cannot establish the ``universality'' of a central
$\rho_{\rm dm}\propto r^{-1}$ structure, even among only cluster-sized halos.
It is worth noting, then, that clusters at $z\simeq0.3$ in the CNOC survey
are also well fit by an \markcite{nav97}NFW model for the dark matter
(\markcite{car97}Carlberg et al.~1997). Moreover, the \markcite{nav97}NFW
concentration of Virgo, $c\equiv r_{200}/r_s=2.8\pm0.7$, is consistent with
that of the CNOC clusters. This $c$ seems low, however, by comparison with
$\Lambda$-CDM simulations of halo evolution (\markcite{nav97}NFW;
\markcite{tho98}Thomas et al.~1998). The physical origin of this
discrepancy---and, for that matter, of the $r^{-1}$ cusp itself---is unclear.

\acknowledgments

Marisa Girardi provided the velocity dispersion data plotted in Figure
\ref{fig2}, and an anonymous referee offered many helpful comments. This
work was supported by NASA through grant number HF-1097.01-97A awarded by the
Space Telescope Science Institute, which is operated by the Association of
Universities for Research in Astronomy, Inc., for NASA under contract
NAS5-26555.

\clearpage

\figcaption[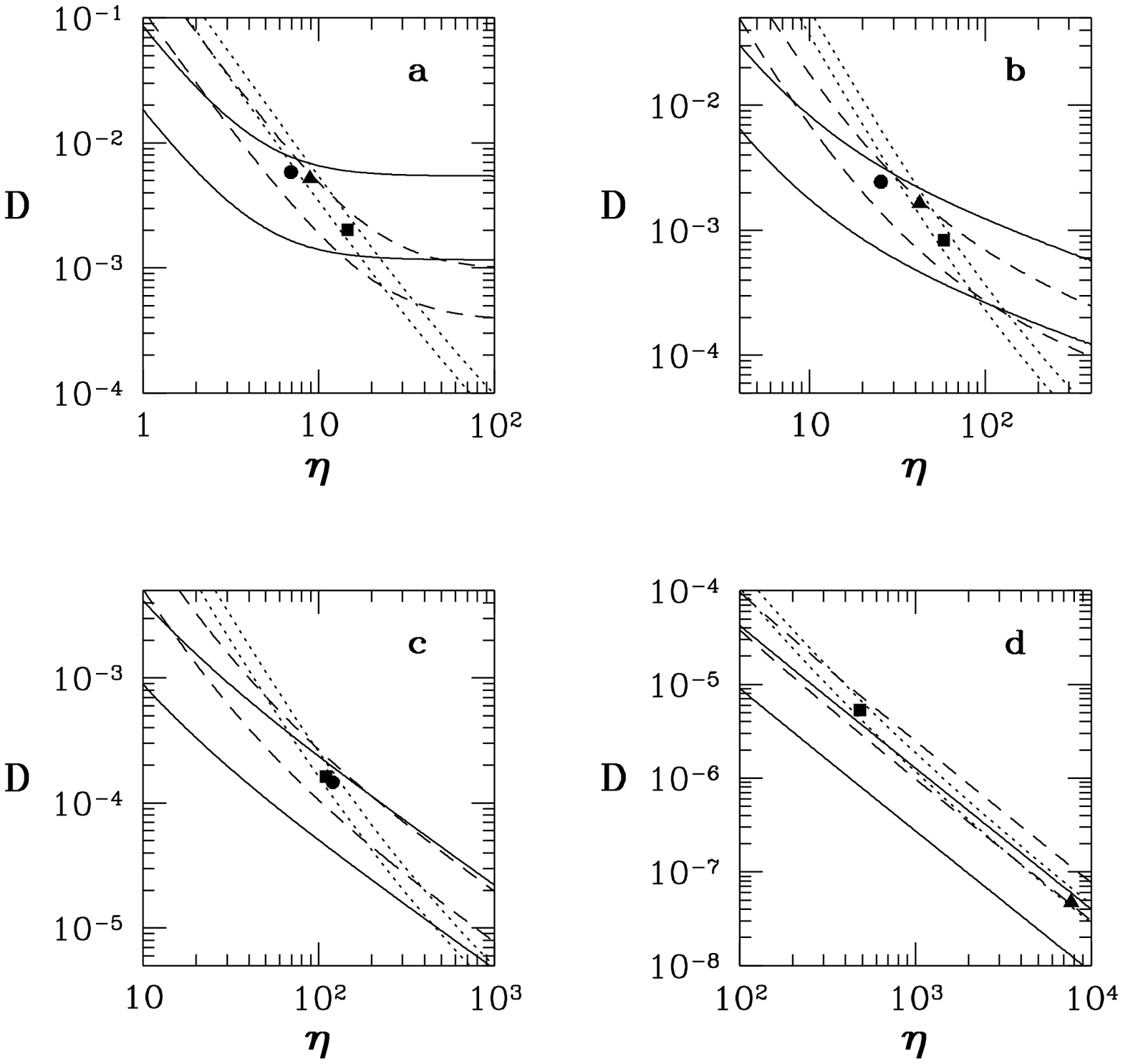]{Tests of four different models for the dark matter
halo in Virgo. In each panel, solid lines contain all $(\eta,D)$ pairs which
(by eq.~[\ref{eq:2}]) satisfy the 2-$\sigma$ {\it limits} on $M_{\rm tot}(30\,
{\rm kpc})$; dashed lines contain those which satisfy the limits on $M_{\rm
tot} (150\,{\rm kpc})$; and dotted lines contain those which (by
eqs.~[\ref{eq:3}]--[\ref{eq:5}]) satisfy the limits on $\sigma_{\rm ap}
(2\,{\rm Mpc})$. Circles in each case mark the single $(\eta,D)$ pair that
yields the {\it best estimates} for both $M_{\rm tot}(30\,{\rm kpc})$ and
$M_{\rm tot} (150\,{\rm kpc})$ [there is no such pair for the steep central
cusp in (d)]; triangles yield the best $M_{\rm tot} (30\,{\rm kpc})$ and the
best $\sigma_{\rm ap}(2\,{\rm Mpc})$; and squares give the best $M_{\rm tot}
(150\,{\rm kpc})$ and $\sigma_{\rm ap}(2\,{\rm Mpc})$.
\label{fig1}}

\figcaption[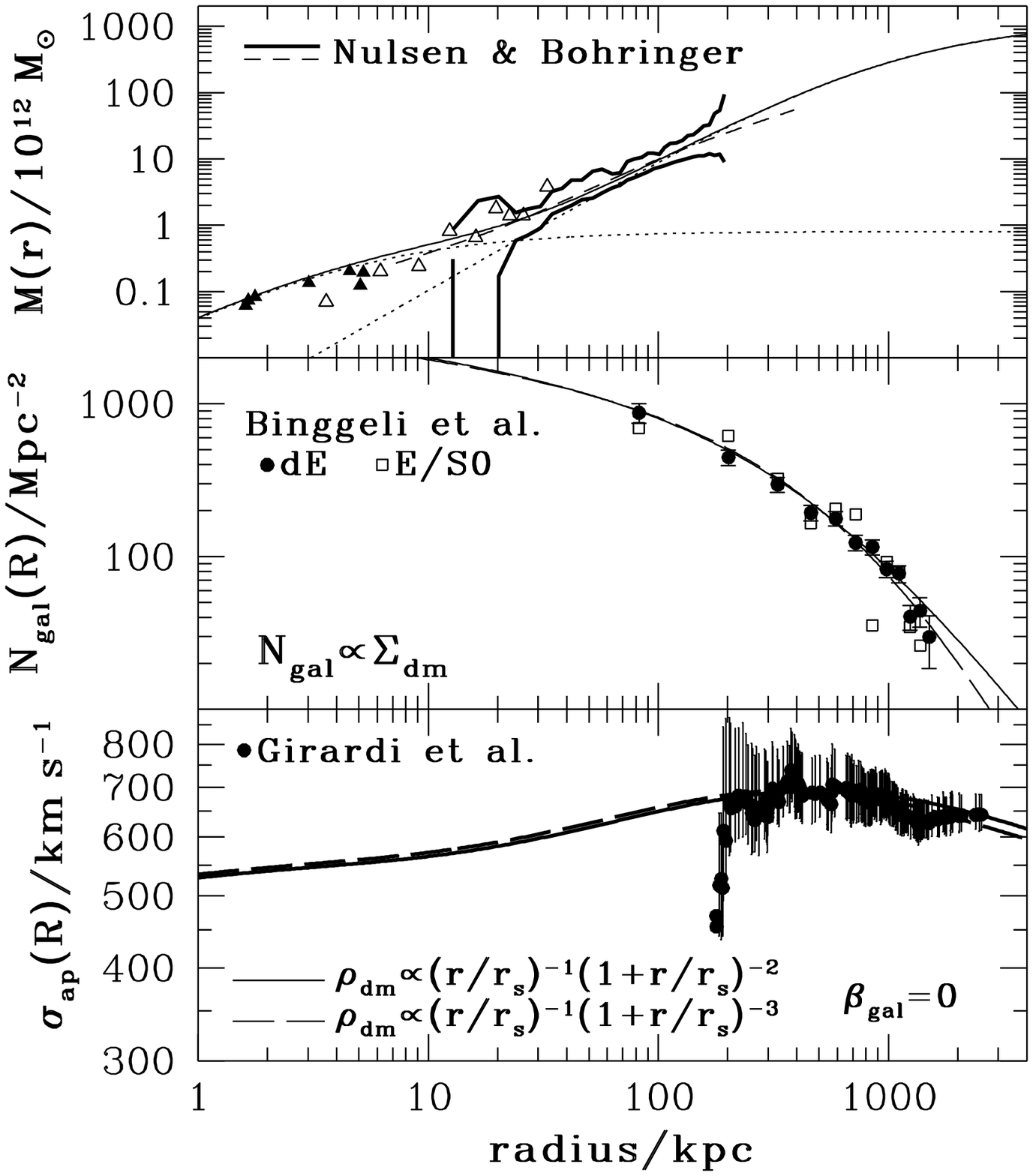]{Comparison of best-fit mass models with the
Virgo data (scaled for an assumed distance of 15 Mpc), for both NFW and
Hernquist (1990) dark matter distributions. Parameters for the NFW halo follow
from panel (c) of Fig.~\ref{fig1}; those for the Hernquist halo come from a
similar procedure. The dotted lines in the top panel trace the separate mass
components $M_{\rm M87}(r)$ and $M_{\rm dm}(r)$. In the middle panel, the
profile of E/S0 galaxies has been scaled to that of the dE's according to the
ratio (697/56) of their total populations in Binggeli et al.~(1987).
\label{fig2}}

\figcaption[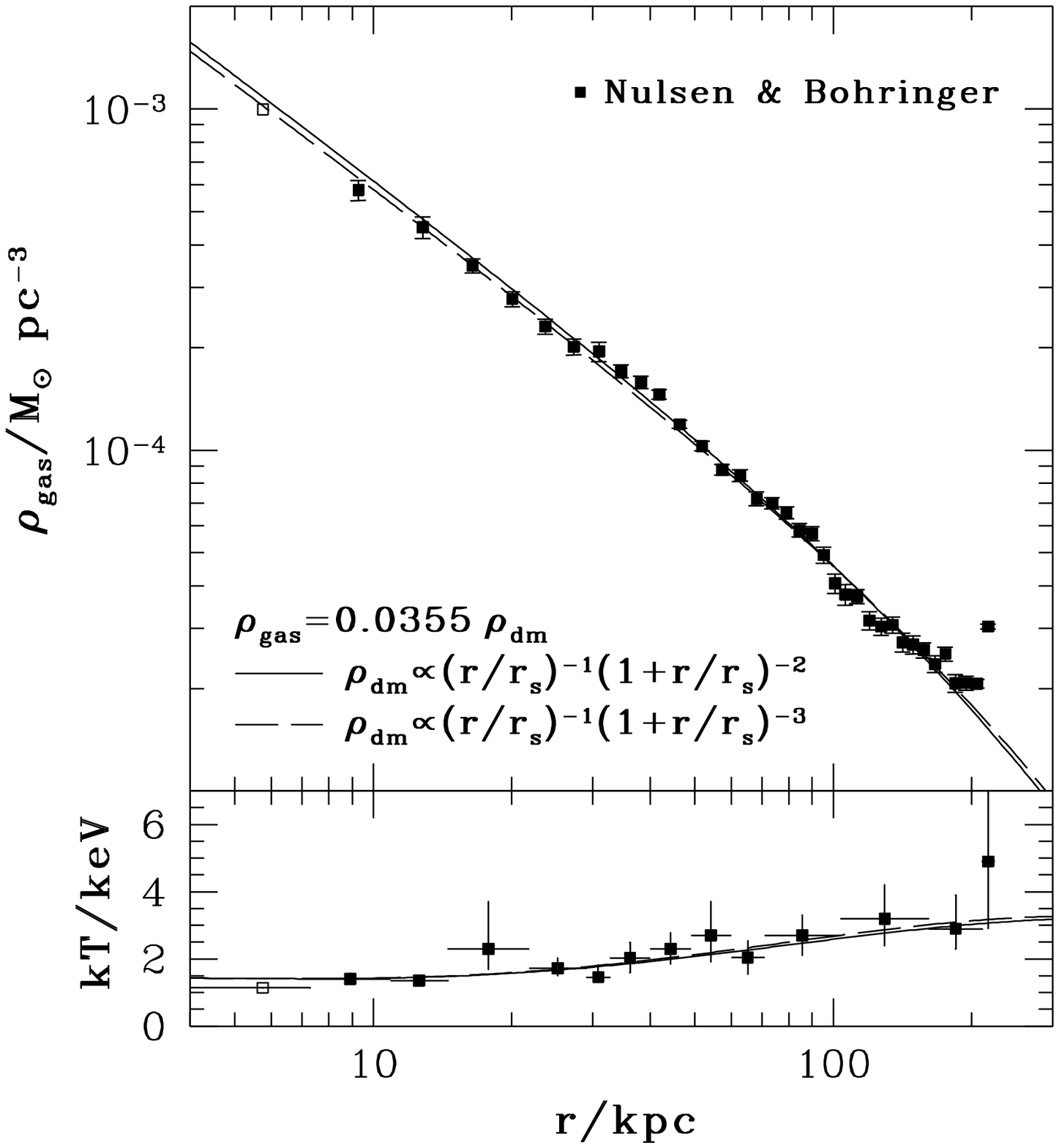]{Comparison of dark matter and intracluster gas
density and temperature profiles. Data and (2-$\sigma$) errobars are those from
rings 2 through 37 of NB95. Points in the innermost
annulus are somewhat uncertain, due to possible contamination by emission from
the center of M87, while densities in the outermost one or two annuli may
suffer from an inaccurate correction for background emission (NB95). Curves in
the bottom panel trace the virial temperature profile of the dark matter halo,
in the total potential arising from $\rho_{\rm M87}+\rho_{\rm dm}$. See NB95
for discussion of the $T(r)$ measurements.
\label{fig3}}

\clearpage

\begin{deluxetable}{ll}
\tablecaption{Mass Model for the Virgo Cluster \label{tab1}}
\tablewidth{0pt}
\startdata
 & \nl
$\rho_{\rm M87}={{3-\gamma}\over{4}}\,
{{\Upsilon_B L_B}\over{\pi a^3}}\,\left({r\over{a}}\right)^{-\gamma}\,
\left(1+{r\over{a}}\right)^{\gamma-4}$ &
 $\gamma=1.33$\ \ \ \ \ \ \ \ $a=5.1\pm0.6$ kpc \nl
$M_{\rm M87}=\Upsilon_B L_B \left({{r/a}\over{1+r/a}}\right)^{3-\gamma}$ &
 $L_B=(5.55\pm0.50)\times10^{10}\,L_{\odot,B}$ \nl
 & $\Upsilon_B=14.6\pm0.2\,M_\odot\,L_{\odot,B}^{-1}$ \nl
 & \nl
$\rho_{\rm dm}=K\,\left({r\over{r_s}}\right)^{-1}\,\left(1+{r\over{r_s}}
\right)^{-2}$ & $K=(3.2^{+2.6}_{-1.3})\times10^{-4}\,M_\odot$\,pc$^{-3}$ \nl
$M_{\rm dm}=4\pi\,K r_s^3\left[\ln\left(1+{r\over{r_s}}\right)-\left({{r/r_s}
\over{1+r/r_s}}\right)\right]\qquad$ &
 $D=\pi a^3 K/\Upsilon_B L_B=(1.65^{+1.35}_{-0.70})\times10^{-4}$ \nl
 & $r_s=560_{+200}^{-150}$ kpc\ \ \ \ $\eta=r_s/a=110_{+40}^{-30}$ \nl
 & \nl
$\rho_{\rm dm}=K\,\left({r\over{r_s}}\right)^{-1}\,\left(1+{r\over{r_s}}
\right)^{-3}$ & $K=(1.6\pm0.7)\times10^{-4}\,M_\odot$\,pc$^{-3}$ \nl
$M_{\rm dm}=2\pi\,K r_s^3\left({{r/r_s}\over{1+r/r_s}}\right)^2$ &
 $D=\pi a^3 K/\Upsilon_B L_B=(8.1\pm3.6)\times10^{-5}$ \nl
 & $r_s=1.07_{+0.36}^{-0.20}$ Mpc\ \ \ \ $\eta=r_s/a=210_{+70}^{-40}$ \nl
\enddata
\end{deluxetable}

\clearpage

\plotone{demfig1.eps}

\plotone{demfig2.eps}

\plotone{demfig3.eps}

\end{document}